Version: 5 December 2023

# A Hedonic Index for Collectables Arising from Modelling Diamond Prices

N I Fisher[1] & A J Lee

## Abstract

This article describes a case study concerned with modelling the price of wholesale diamonds, as part of a project to develop an online diamond auction platform. The work was extended to exploring how to develop an index that could be used to track market trends of wholesale diamond prices. The approach we used is readily generalised to defining market indices for so-called "collectables, and can provide the basis for construction of derivatives. With the burgeoning interest in new markets of collectables such as those generated by the concept of a Non-Fungible Token, it is reasonable to suppose that there will be concomitant increasing interest in developing derivatives for these markets.

## Key words

Derivative, Forecasting, Hedonic Collectable Index, Non-fungible Token

## 1. INTRODUCTION

This article has its origins in a very simple question: *How much does a diamond cost*? The question arises if, for example, one wishes to launch an online venture to disrupt the diamond auction market, as was indeed the case here. And, as is often the case, the answer starts "It depends … ".

---

[1] Nicholas Fisher is Visiting Professor of Statistics, University of Sydney, and Principal of ValueMetrics Australia. Alan Lee is Emeritus Professor of Statistics, University of Auckland, New Zealand.
Corresponding author: Professor Nicholas Fisher, School of Mathematics & Statistics, University of Sydney, NSW 2006 AUSTRALIA. Email Nicholas.Fisher@sydney.edu.au

To begin with, a "gem-quality" diamond[2] has eight distinguishing characteristics, each of which has several (in most cases many) levels[3]. The most important of these are the so-called 4 C's, Carat, Colour, Clarity and Cut; however, Shape, Polish, Symmetry and Fluorescence can also affect the price significantly. For each characteristic, the differing levels provide an ordering in terms of the quality of the diamond. Thus for a given weight (Carat size) of diamond, the top quality diamond is *Round*, with Colour *D*, Clarity *Flawless*, Cut *Excellent*, Polish *Excellent*, Symmetry *Excellent* and Fluorescence *None*. We shall refer to such a diamond as a "perfect" diamond. Location of the sale and whether it a wholesale or retail purchase also affect the price, not to mention factors such as the cachet associated with a small number of famous diamonds, and whether it is a natural or laboratory-grown diamond[4].

And finally, the pricing of a diamond is, in many cases, more artful than scientific. Large wholesale diamond merchants publish regular wholesale price lists although the formulae for producing these lists are proprietary. However, for the 250,000 or so retail jewellery shops around the world, the pricing of a diamond is, in the words of one expert, derived from "guesswork, some usage of wholesale diamond pricing, advisors and consultants and the direction of the wind."

The online diamond venture was conceived as having a number of aspects apart from online auctions. These included providing valuations for insurance purposes and, in the longer term, the production of regular diamond market indices and associated products that are the main focus of this article. As a prelude, we sketch our approach to estimating the price of a diamond.

**Wholesale diamonds** A lot of data are available (*e.g.* IDEX 2023, Rapaport 2023, VDB 2023). For example, subscribers to IDEX are able to purchase downloads of current wholesale prices for some 1M diamonds. The modelling we conducted was based on weekly downloads extending back to 2015. Several nonparametric and semi-parametric approaches to modelling the wholesale price of GIA-certified diamonds[5] as a function of the eight diamond characteristics plus Location and Date were explored. The two best-performing methods, as judged by relative percentage error of prediction, were Random Forests and Gradient Boosting, with the former preferred because it was simpler to capture the model in an app suitable for a mobile device. The model was subsequently tested and found appropriate for corresponding data from another provider.

---

[2] "Gem-quality" refers to diamonds with minimal impurities and defects. They have a specific gravity very close to 3.52. So-called "industrial" diamonds are of lower quality, and the only meaningful attribute affecting their price is size.

[3] See *e.g.* https://www.gia.edu/diamond-quality-factor.

[4] The market for so-called lab-grown diamonds (LGDs) is quite different from that of natural diamonds, but also changing rapidly. *See e.g.* https://www.smh.com.au/national/lab-grown-diamonds-are-marketed-as-the-ethical-choice-are-they-really-20230309-p5cqq5.html.

(It is appropriate to make explain why laboratory-grown "gem-quality" diamonds, which are very difficult to distinguish from natural diamonds, are not considered further, in this article. When new, a laboratory-grown diamond is priced at about 50% of the corresponding natural diamond. However, it retains virtually none of its original value, as a second-hand diamond.)

**Auction sales diamonds** Only small numbers of diamonds are auctioned annually, and these data are not readily obtained. We have access to some 2500 individual diamond sale prices dating back to 2015, with just the 4 C's available for each. The data set is too small for the nonparametric approach and we finally settled on a generalised linear model for log(Sales Price).

## 2. CONSTRUCTING INDICES FOR DIAMONDS AND OTHER COLLECTABLES

In devising a useful market index for diamonds, it becomes evident that the same issues might arise more generally for a wide range of objects generally known as Collectables (Wikipedia 2023a), where:

> A **collectable** (or **collectible** or **collector's item**) is any object regarded as being of value or interest to a collector. Collectable items are not necessarily monetarily valuable or uncommon. There are numerous types of collectables and terms to denote those types. An antique is a collectable that is old. A **curio** is a small, usually fascinating or unusual item sought by collectors. A **manufactured collectable** is an item made specifically for people to collect.

Thus, the sorts of objects that people might collect and perceive to be of value to others include precious stones, vintage cars, antique furniture, coins, stamps, comics, fine wine, old spirits (whisky, cognac, …), watches, teaspoons, and so on. Some collectable markets are very active (diamonds, old comics, vintage cars) and others erratic at best (old single malt whiskies).

---

[5] The Gemological Institute of America (https://www.gia.edu/) provides a certification service for diamonds and some other precious stones.

For an active market (e.*g*. Koehn 2023; Lukpat 2023) it may be of interest to be able to gauge overall movements in the market, either for direct trading purposes or as a basis for associated investment instruments using index futures, a purpose served by an appropriate *market index*.

Desirable characteristics of such an index would include that it derive from readily available data; respond sensibly to new sales or price quotations; respond sensibly in terms of capturing trends and volatility, and differentially to the varying characteristics of a given collectable; and allow for retrospective calculation, to produce a historical series.

Of course, market indices are of interest not just in their own right, as a means of capturing changes in a given market. They also provide the basis for constructing financial instruments known as "derivatives[6]". With the burgeoning interest in new markets such as those generated by the concept of a Non-Fungible Token (NFT), it is reasonable to suppose that there will be concomitant increasing interest in developing derivatives for these markets.

Our specific focus will be on the wholesale diamond market. However, the approach that we describe appears to be applicable to any collectable for which adequate market data are available on an ongoing basis.

## 3. DEVELOPING A MARKET INDEX FOR A COLLECTABLE ITEM

We shall describe our approach using (natural) diamonds as an example because the market is sufficiently complex as to rule out a simple solution.

A seemingly straightforward approach to defining an index is to select the price of a 1ct (one Carat) perfect diamond, traded in New York. However, there are a number of problems with such a definition. For example,

(a) there are relatively few perfect diamonds of exact size 1ct, and so virtually no trading activity at any given time.

(b) 1ct represents an important "boundary value" because of the cachet of being "at least 1ct". There is a far greater jump in price from 0.99ct to 1ct, than from 1ct to 1.01ct, for any combination of the other characteristics.

---

[6] From Investopedia: "A derivative is a contract between two or more parties whose value is based on an agreed-upon underlying financial asset (like a security) or set of assets (like an index). Common underlying instruments include bonds, commodities, currencies, interest rates, market indexes, and stocks."

Version: 5 December 2023

   (c) trade in diamonds is very responsive to fashion. For example, at any given time there might be temporary heightened interest in pink diamonds, or in cushion-shaped diamonds as worn by a prominent public figure at her wedding, and so relative inaction in the round white diamond sector of the market.

and so on. Thus this sort of simple approach does not produce an index that satisfies the desirable characteristics described earlier. Additionally, it does not lend itself to providing a generic approach that might be adapted readily to other collectables.

As an alternative, one can ask: *What approaches have been used in other markets that might be qualitatively similar to the diamond market?* The seemingly remote case of real estate furnishes such an analogy.

The price of a house depends on a variety of factors, some of which are important considerations (Location, Size, Proximity of shops and transport, …) and others less so. The importance of house price indexes has been discussed, for example, by Hill (2013); this article also explains why so-called *hedonic indexes* are replacing simpler, more traditional indices, as a guide to the behaviour of the housing market. Investopedia (Investopedia 2023) provides a concise description of hedonic pricing, which it traces back to Rosen (1974). Accordingly, we sought to develop hedonic indexes for collectables where in the language of Wikipedia (Wikipedia 2023b), a hedonic index is

> "Hedonic pricing is a model that identifies price factors according to the premise that price is determined both by internal characteristics of the good being sold and external factors affecting it."

Accordingly, we sought to develop hedonic indexes for collectables where, in the language of Wikipedia (Wikipedia (2023b)), a hedonic index is
> "… any price index which uses information from hedonic regression, which describes how product price could be explained by the product's characteristics."

A key element in the definition of our proposed generic Hedonic Collectable Index (HCI) is that a collectable in any given category (*e.g.* diamonds) can be categorized into groups within each of which they are somewhat homogeneous with respect to their values. We shall exemplify this with diamonds.

Suppose that at time $t$, we have available a set of $N$ diamonds $D_1, \ldots, D_N$, with diamond $D_k$ having diamond characteristics $x_k = \{$Carat size, Colour, $\ldots\}$ and associated (sales or wholesale) price $P_k(t)$. Also, let $F_k(t)$ be the prediction of the price of $D_k$, made at time $t$, based on a model fitted to data up to some fixed prior time $t_0$, referred to the baseline. The price index is based on comparing the actual price with the predicted price for each diamond using the ratio $R_k(t) = P_k(t) / F_k(t)$.

The rationale for this definition is as follows. If the relationship between the price of a diamond and the various diamond characteristics does not change between time $t_0$ (when the predictive model is fitted) and time $t$, the predictions using this model made at time $t$ should be unbiased, and the average ratios $R_k(t)$ should have an average value very close to unity. We are assuming that the predictor is also unbiased for data collected at time $t_0$ (*i.e.* the data used to fit the prediction model) so the average ratio should not change appreciably between times $t_0$ and $t$. Conversely, if the price level increases between time $t_0$ and $t$, then the predictions (using the predictor based on the snapshot at time $t_0$) made at time t will tend to under-predict the prices and the average value of the ratios $R_k(t)$ will be more than unity. And conversely, if prices go down, then the model will over-predict and the ratios will have an average value of less than unity. In practice, the simple ratios $R_k(t)$ are too volatile to be used for an index and need some form of smoothing before use.

## 4. A HEDONIC COLLECTABLE INDEX (HCI) FOR WHOLESALE DIAMOND PRICES

As noted earlier, large volumes of wholesale price data for diamonds are readily available. We shall show how the methodology can be applied to a data base currently comprising about one million diamonds, with prices updated weekly. (In separate studies to date, there appear to be no differences in price structure from source to source.) We shall use the term 'snapshot' to refer to a (large) set of wholesale diamond data (characteristics and associated prices) for any particular week.

Let $G_1, \ldots, G_M$ denote a partitioning of the set of all possible diamonds into a number of mutually exclusive groups according to some concept of the *weight* of a diamond (to be clarified shortly). Associated with these groups are weighting factors $W_1, \ldots, W_M$, $W_1 + \ldots + W_M = 1$, that indicate the relative importance of each of these groups. Then each $D_k$ will fall into precisely one of these groups ($G_r$, say) and so is assigned the weighting factor $w_k = W_r$.

A new value of the wholesale index is produced for each snapshot. For each diamond in the snapshot, the ratio $R_k(t)$ is calculated, and the ratios averaged over each group, giving $M$ group averages $\overline{R_k(t)}$. The HCI is a weighted average of these $M$ averages using the associated group weighting factors $W_k$:

$$\mathrm{HCI} = \{W_1 \overline{R_1(t)} + \ldots + W_N \overline{R_M(t)}\}$$

In this example, we have used snapshots dating back to 15 January 2015, so the hedonic index can be set up as follows:

(a) use only diamonds in the range 0.25 ct ≤ Weight < 100 ct. (There are too few wholesale prices for diamonds outside this range to model Price satisfactorily.)

(b) the predicted diamond prices in the HCI are derived from the model fitted to the initial 15 January 2015 snapshot.

Figure 1 shows the HCI for wholesale prices for the period 1 January 2020 – 30 August 2022.

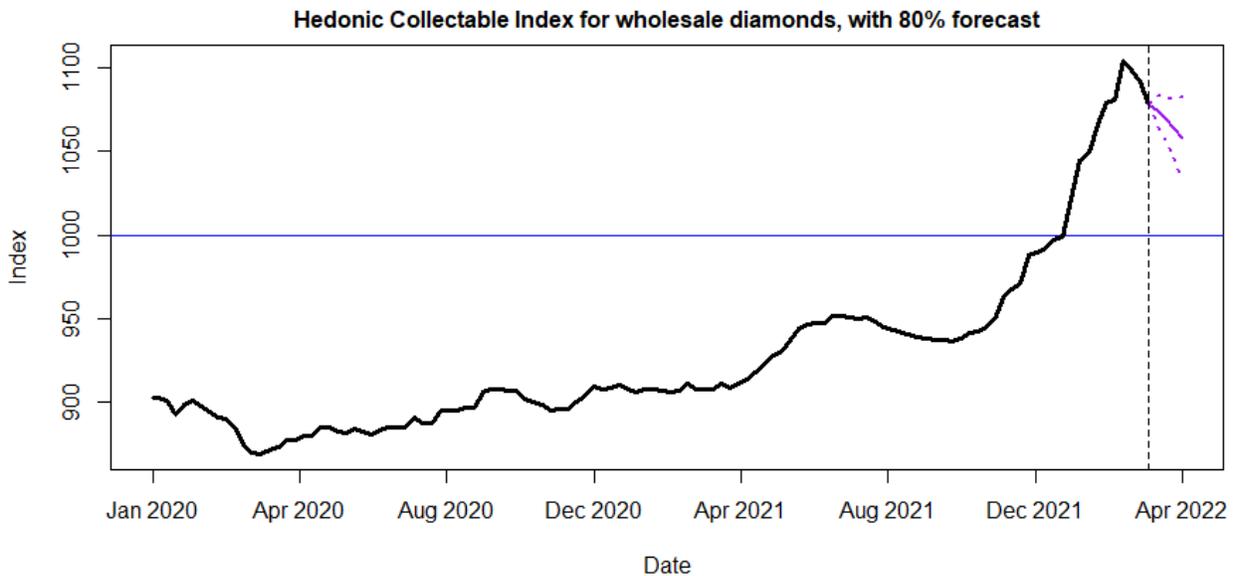

Figure 1. Hedonic Collectable Index for diamond wholesale price data, January 2020 – April 2022, showing forecast values of HCI with an 80% forecast interval.

## 5. CHOICE OF GROUPS AND WEIGHTS

The price of diamonds has a threshold effect at certain discrete carat sizes. For example, there is a considerable jump in price when a diamond crosses the one-carat threshold. There is a similar "price bump" at 0.5, 2,3,4,5 carats. Accordingly we divide the diamonds in a snapshot into seven classes 0–0.49 carats, 0.50–0.99 carats, 1.00–1.99 carats, 2.00–2.99 carats, 2.00–2.99 carats, 3.00–3.99 carats, 4.00–4.99 carats and 5+ carats. In a typical snapshot, there are many more diamonds in the smaller classes: for example for the snapshot of 27/06/2022 the class sizes are shown in Table 1.

| Diamond weight (carats) | | | | | | |
|---|---|---|---|---|---|---|
| **0-0.49** | **0.50-0.99** | **1.00-1.99** | **2.00-2.99** | **3.00-3.99** | **4.00-4.99** | **5+** |
| 329506 | 281696 | 134001 | 24661 | 8710 | 2333 | 4937 |

Table 1. Class sizes for the 27/06/2022 snapshot.

While the choice of weights is somewhat arbitrary, one reasonable proposal is to choose weights proportional to the total value of the diamonds in a group. This is the analogue of market capitalization in stock indices. For the snapshot depicted in Table 1, the total value of each group is shown in Table 2.

| Total value of group (millions of dollars) | | | | | | |
|---|---|---|---|---|---|---|
| **0-0.49** | **0.50-0.99** | **1.00-1.99** | **2.00-2.99** | **3.00-3.99** | **4.00-4.99** | **5+** |
| 120 | 526 | 1136 | 591 | 419 | 157 | 813 |

Table 2. Total value of diamonds in the seven groups for the 27/06/2022 snapshot.

If we choose weights proportional to total value, changes in the value of large diamonds ($> 1ct$) will tend to dominate the index, even though there are many more small diamonds ($< 1ct$). Thus, the index will tend to mask changes in the value of small diamonds. This can be reduced to some extent by combining the "market capitalization" weights with a set of equal weights, putting weight 1/7 on each group. This leads to a set of 'final' weights

$$w_f = \frac{w_p + 1/7}{2}$$

where $w_p$ represents the "proportional" weights. This leads to a final set of weights shown in Table 3.

| Final weights | | | | | | |
|---|---|---|---|---|---|---|
| **0-0.49** | **0.50-0.99** | **1.00-1.99** | **2.00-2.99** | **3.00-3.99** | **4.00-4.99** | **5+** |
| 0.098 | 0.141 | 0.222 | 0.150 | 0.127 | 0.092 | 0.180 |

Table 3. Final weights for the seven groups for the 27/06/2022 snapshot.

The distribution of the ratios $R_k(t)$ is somewhat right-skewed, with a few relatively large ratios. This suggests using medians instead of means to characterise the centres of these distributions. However, the sample sizes in our data are so large that the effect of the large outliers is swamped by the bulk of the data, and the use of medians makes little difference.

# 6. BEHAVIOUR OF THE HCI IN RESPONSE TO SUDDEN MARKET SHIFTS

We present two examples of how the HCI responds to abrupt changes in the market.

**Scenario 1: Change of fashion.** Immediately after a Royal engagement has been announced, an image of a Cushion-shaped engagement ring flashes around the world. This has two effects. For a period of 3 months, there is a 5% decline in sale of Rounds, which change to Cushions. And there is a 5% increase in the price of all Cushions.

**Scenario 2: Slump in prices of small diamonds (< 1ct).** Over a period of 3 months (13 snapshots), for the first 3 snapshots, there is a 5% decline in price, followed by a 10% decline for the next 7 snapshots and then a 5% decline for the last 3 snapshots.

Figure 2 shows that each of these scenarios has a clear impact on the HCI.

# 7. DISCUSSION

## 7.1.    Use of sub-indices

We noted above that the choice of weights was somewhat arbitrary. The current choice still gives more weight to the larger diamonds and may mask price movements in the small diamonds if these differ from movements on the price of the larger diamonds. In fact, in the current snapshot, the price of small diamonds has hardly moved from the baseline while the larger diamonds have increased in price, thus increasing the ratios $R_k(t)$. The use of a single index, no matter how weighted, will not capture these sorts of differential price movements. A simple remedy is to augment the main index with a series of sub-indices, consisting of the mean ratios for each group. In fact, we may define sub-indices for a variety of groupings, not just those based on carat weight. We could, for example, consider sub-indices based on Colour or Shape.

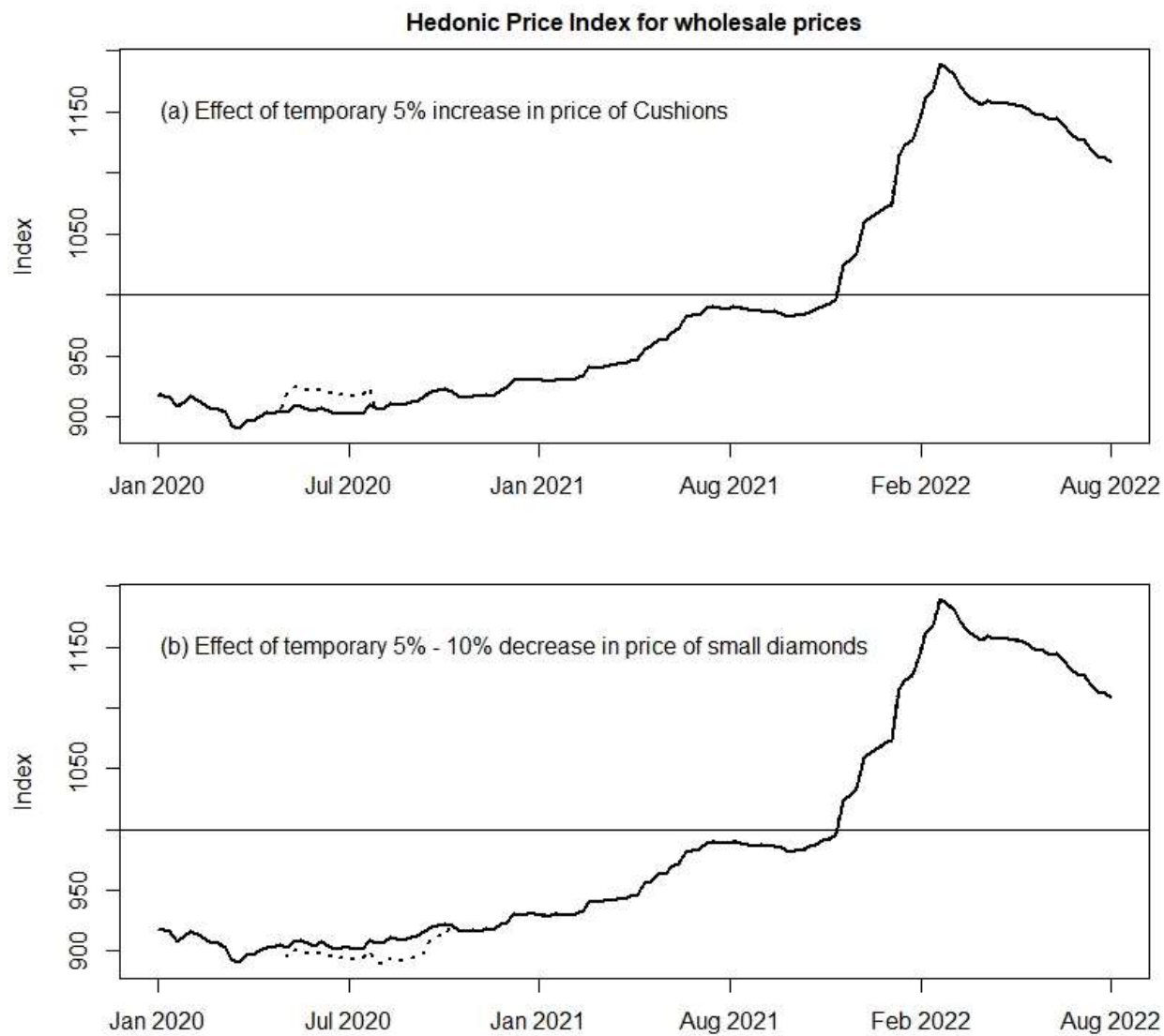

Figure 2. Effect of sudden market changes on the HCI, as indicated by the dashed segments in each graph. (a) Over a period of 3 months, a sudden 5% increase in the price and volume of Cushion-shaped diamonds, with a corresponding 5% decrease in sales of Round diamonds. (a) Over a period of 3 months, a sudden 5% increase in the price and volume of Cushion-shaped diamonds, with a corresponding 5% decrease in sales of Round diamonds. (b) Over a 3-month period, a general decline in overall prices (5% / 10% / 5%).

## 7.2. Comparison with other diamond-specific indices, and extensions

There are no non-proprietary wholesale diamond indices currently available that cater for the full spectrum of wholesale diamonds.

Detailed and lengthy series data for leading international market indices are freely available from several websites (*e.g.* Investing.com (2023)), providing a means of evaluating the behaviour of Hedonic Collectable Indices. As an example, Figure 3 shows how the HCI compares with several indices. (The other market indices have been rescaled and re-centred so that the relative patterns of movement can be assessed.)

### 7.3. Forecasting the HCI

It is possible to forecast the HCI using standard forecasting methods. In Figure 1, the series is forecast four weeks ahead using a Holt-Winters forecast (see *e.g.* Brockwell & Davis (2002)), with the dotted lines representing the 80% error bounds on the forecast. As can be seen, the error bounds are quite wide. An alternative approach is to fit an ARIMA model to the series. In the present example, this resulted in a forecast with slightly wider error bounds. In fact, it is possible to improve forecasting by utilising information from some other index series, as a means of exploring the possibility of forecasting notional HCI Futures.

### 7.4. Nature of the data

The wholesale diamond example is based on large amounts of price data being available on a regular basis (here, weekly). It is readily modified for other intervals (daily, monthly, ...). Data deriving from sales are necessarily different in type. For a collectable market operating internationally, a sale (even multiple sales) can occur at any time in a 24-hour period. However, the principle for constructing a Hedonic Collectable Index is the same: it is based on the ratios of the actual price of a collectable to the price predicted by a model fitted to previous prices.

### 7.5. Updating the index

After a number of years, it may be appropriate to update the weighting factors owing to the impact of inflation on the Price classes. The level of the newly-calculated HCI then needs to be calibrated to provide a smooth transition from the earlier values, and a change-of-definition time-point noted in any published series.

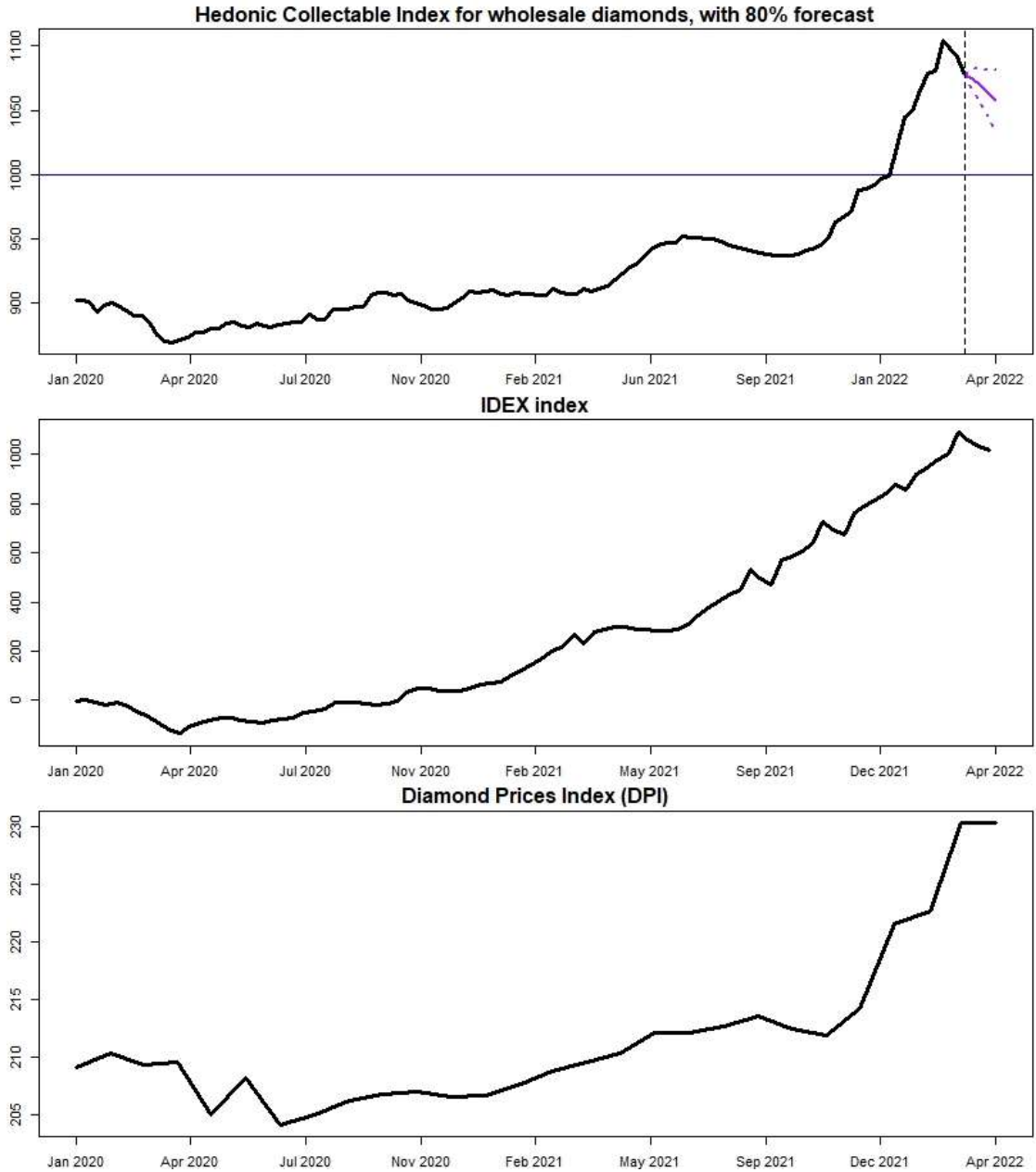

Figure 3. A comparison of the Hedonic Collectable index with three other indices, for publicly available periods. There is moderate agreement in terms of trends.

## 7.6. Confidence intervals

If we regard the snapshot as a random sample from a larger population, the HCI described above is an estimate of the corresponding population index

$$HCI_{POP} = \sum_{g=1}^{G} w_g E[R_k(t)| (P_k(t), x_k(t)) \in G_g]$$

To develop an expression for the standard error of this estimate we make the standard error conditional on the baseline data set. Thus we regard the baseline predictor and the weights as given and fixed. We can take advantage of the special structure of the index as a weighted average of means. The variance of the index is

$$Var(HCI) = \sum_{g=1}^{G} w_g Var[\overline{R_k(t)}| (P_k(t), x_k(t)) \in G_g]$$

We could then estimate each component variance separately: since each component is the variance of a mean, we can estimate it by the usual formula *Var* (sample mean) = (Variance / sample size).

Various confidence intervals can be calculated. One possibility is to use the standard error suggested above for a standard normal theory confidence interval. Alternatively, we can use a bootstrap confidence interval based on the quantiles of the bootstrapped HCI values or a percentile-*t* interval. Again, for the data set used earlier to compare estimated standard errors, the three intervals are similar, as shown in Table 2, and the extra computation of the bootstrap is not necessary.

|  | 2.5% | 97.5% |
|---:|---|---|
| Normal | 1166.389 | 1167.658 |
| Bootstrap | 1166.591 | 1167.490 |
| Percentile-t | 1166.556 | 1167.456 |

Table 2 . Comparison of three methods of calculating confidence intervals for HCI

## Acknowledgements

We are grateful for financial support from Fine Art Bourse Australia Pty Ltd and the Australian Federal Government Research & Development Scheme while carrying out this research.

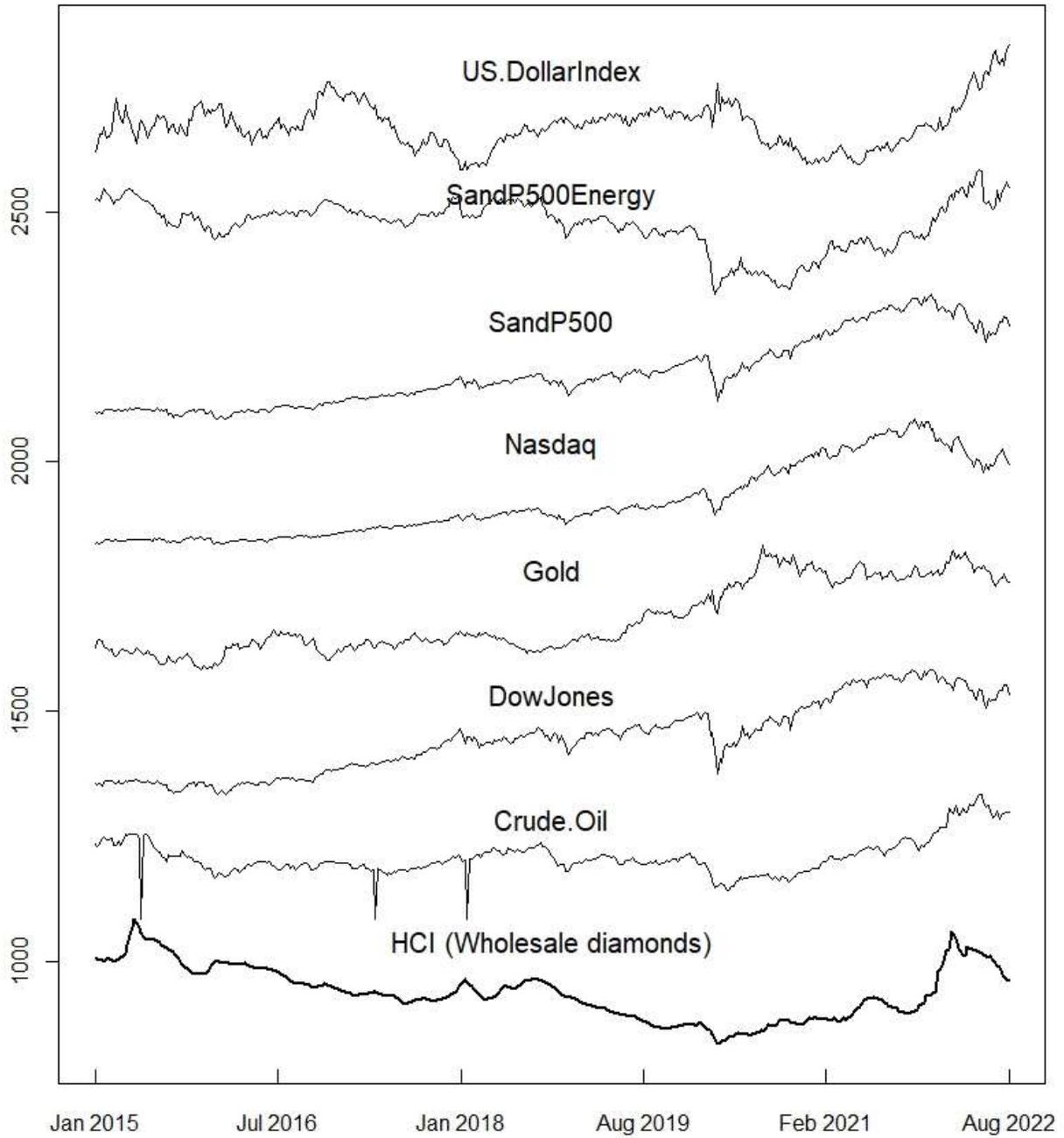

Figure 4. Comparison of the wholesale diamond HCI with some standard market indices, for the period January 2015 - August 2022. All indices except HCI have been transformed (re-centred and re-scaled) for purposes of comparison with HCI.

Version: 5 December 2023